\documentclass{aastex62}

\graphicspath{{./}{figures/}}

\received{}
\revised{}
\accepted{}
\submitjournal{ApJ}

\shorttitle{Sample article}
\shortauthors{Cheng et al.}
\begin{document}

%\title{Near-infrared variability among stars in a distant young protocluster}
\title{Stellar variability  in a forming massive star cluster}

\correspondingauthor{Yu Cheng}
\email{ycheng.astro@gmail.com}

\author[0000-0002-0786-7307]{Yu Cheng}
\affil{Dept. of Astronomy, University of Virginia, Charlottesville, Virginia 22904, USA}

\author{Morten Andersen}
\affiliation{Gemini Observatory, NSF’s National Optical-Infrared Astronomy Research Laboratory Casilla 603, La Serena, Chile}

\author{Jonathan Tan}
\affiliation{Dept. of Space, Earth \& Environment, Chalmers University of Technology, Gothenburg, Sweden}
\affil{Dept. of Astronomy, University of Virginia, Charlottesville, Virginia 22904, USA}



\begin{abstract}
We present a near-infrared (NIR) variability analysis for an 6\arcmin
$\times$ 6\arcmin~ region, which encompasses the massive protocluster
G286.21+0.17. The total sample comprises more than 5000 objects, of
which 562 show signs of a circumstellar disk based on their infrared colors.
The data includes HST observations taken in two epochs separated by 3
years in the F110W and F160W bands. 363 objects (7\% of the
sample) exhibit NIR variability at a significant level (Stetson index
$>1.7$), and a higher variability fraction (14\%) is found for the young
stellar objects (YSOs) with disk excesses.
% MA: Since the specific fraction is given for the whole sample it should also be done here I think. 
We identified 4 high amplitude ($>0.6$ mag) variables seen in both NIR
bands.
%MA I'd prefer high amplitude variables instead of highly variable 
Follow up and archival observations of the most variable object in
this survey (G286.2032+0.1740) reveal a rising light curve over 8
years from 2011 to 2019, with a K band brightening of 3.5 mag. Overall
the temporal behavior of G286.2032+0.1740 resembles that of typical FU
Ori objects, however its pre-burst luminosity indicates it has a very
low mass ($<0.12\:M_\odot$), making it an extreme case of an outburst
event that is still ongoing.
%Further monitoring is highly encouraged.
%MA concerning the mass. I've used the Baraffe 15 isochrone for 0.5, 1, and 2 Myr. The main problem was that I got widely different results depending if I used JH, HK, and JK as the color-magnitude combination. I then tried with the one by Hosek et al. That helped quite a lot in that the difference is now < 50% (which is still huge, and suggests it's not "normal" reddening alone I think. 
%Anyway, the different mass estimates are: 
%Baraffe 15 isochrone, 1 Myr
%    0.0910424 (for all three cases I used J-H,J first then J-K,J and then H-K,H, respectively  
%    0.0624947
%    0.0515436
%Baraffe 15 isochrone, 0.5 Myr
%    0.0779587
%    0.0478992
%    0.0403584
%Baraffe 15 isochrone, 2 Myr
%     0.104738
%    0.0667793
%    0.0568460

%YC- Even though the mass is not well constrained, the upper limit here ~0.1 Msun is still (one of?) the lowest mass outbursting YSO to my knowledge.
\end{abstract}


%%, repeat ;) 


\section{Introduction}\label{sec:intro}

Variability is ubiquitous among young stellar objects (YSOs).
A low level of variability (i.e., typically below a few 0.1 mag) has been observed in most YSOs in the optical and NIR \citep[e.g.,][]{Parihar09}. Mechanisms to produce such variations include rotationally modulated cool spots, hot spots on the stellar surface, extinction changes, and changes in the inner circumstellar disk \citep{Wolk13}. Some of these mechanisms, like hot spots and varying extinction, may also produce variability with larger amplitudes \citep[see, e.g., ][]{Grankin07,Bouvier13}.
%A low level of variability (i.e., below a few 0.1 mag)
%has been observed in most YSOs in the optical and NIR
%\citep[e.g.,][]{Parihar09}, which could arise from various mechanisms,
%including rotationally modulated cool spots, hot spots on the stellar
%surface, extinction changes, and changes in the inner %\textbf{circumstellar}
%disk \citep{Scholz09}. 
Apart from these common causes of variability,
a small fraction of YSOs
show evidence for eruptive behavior, with variations larger than 1
magnitude in the optical or NIR bands over a few years or
decades. This type of variability is thought to be related to the
process of accretion from the circumstellar disk on to the
protostar. During these bursts the YSO may increase its mass accretion
rate by several orders of magnitude compared with quiescent phases,
resulting in strong variability. While this episodic accretion
scenario is well established, the driving force of this phenomenon
is still poorly understood \citep[e.g.,][]{Audard14}. Understanding
the underlying mechanisms is crucial not only for building a complete
picture of star formation, but also for the potential implications on
the planet formation process \citep[e.g.,][]{Zhu09}.

The nature of YSOs favors observations at near and mid-IR wavelengths,
which allow for direct detection of optically thick disks, e.g., via
excess K-band flux \citep{Lada92}.
%Sort of true but there is quite some work done on the UV excess from YSOs as well. It's just hard to do observationally so there are limited targets. 
% 
Over recent years there has been an increasing interest to search for
eruptive variables with long-term NIR observations. \citet{Scholz12}
used archival NIR photometry to investigate the long-term variability
in a few nearby low-mass star-forming regions and found a low fraction
($\sim$2\% in the YSO sample) of large amplitude variable objects. A higher
incidence of K band variations $>1$ mag ($\sim13\pm$7\%) has been
reported in Class I YSOs in the dark cloud L1003 in Cygnus OB7
\citep{Rice12,Wolk13}. A panoramic search by the UKIRT Infrared Deep
Sky Survey \citep[UKIDSS;][]{Lawrence07} found a strong concentration
of high-amplitude IR variables towards star-forming regions
\citep{Pena14}, and this is confirmed by recent VVV survey (VISTA
Variables in the Via Lactea; \citet{Minniti10}), in which more than
100 eruptive YSOs are detected \citep{Pena17}.

G286.21+0.17 (hereafter G286) is a massive ($\sim2000\:M_\odot$)
protocluster associated with the $\eta$ Car giant molecular cloud at a
distance of 2.5$\pm$0.3 kpc \citep{Barnes10}. The gas and dust
component is well studied with ALMA, which reveals $\sim80$ dense
cores in millimeter continuum emission \citep{Cheng18}. NIR
observations reveal a high disk fraction of the YSOs, which suggests
the cluster is very young \citep[$\sim$ 1~Myr;][]{Andersen17}. Here we
present analysis of two-epoch HST NIR imaging of G286, with the main
goal of characterizing variabilities. In particular, we report the
discovery of a strong outburst in a low-mass embedded YSO, as well as
%its photometric follow-up using VLT and Gemini observations.
its photometric and spectroscopic follow-up using Gemini observations.
 %MA It's a bit misleading to say follow-up with the VLT data since they were obtained before the HST data
\section{Data}\label{sec:style}

\subsection{HST WFC3/IR imaging}

\begin{figure}[ht!]
\epsscale{0.8}\plotone{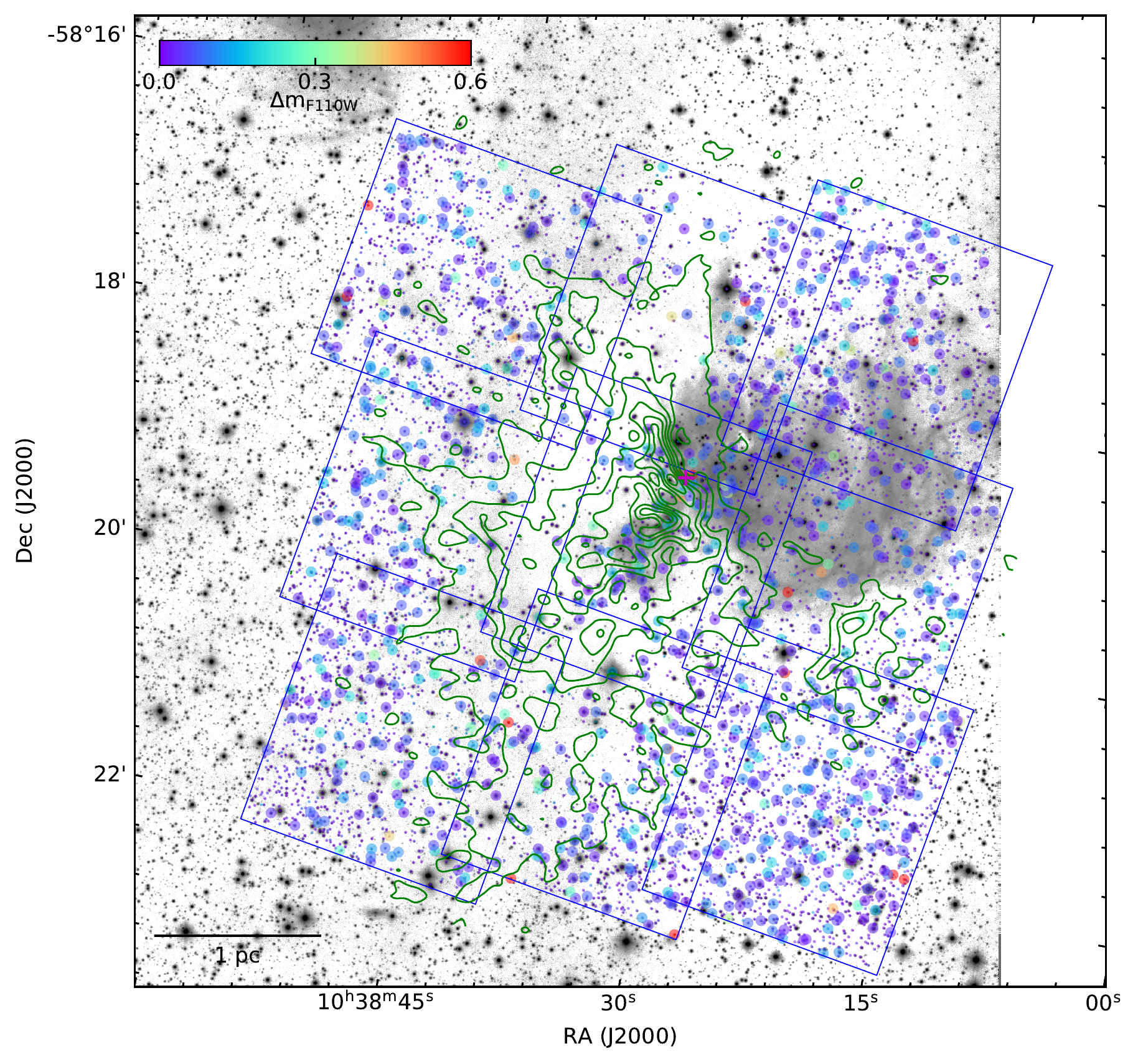}
\caption{
VLT HAWK-I $Ks$-band image of G286 in grey scale. Overplotted in green
contours is ALMA C$^{18}$O(2--1) integration map. 
The contours start from 4 Jy
beam$^{-1}$km s$^{-1}$ in steps of 4 Jy beam$^{-1}$km s$^{-1}$. The
colored circles show the sources detected in the VLT observations,
with the color indicating the F110W band magnitude differences between
two HST epochs. The variables with Stetson index larger than 1.7 are
shown with larger circles. The blue rectangles show the extent of the
3$\times$3 mosaic of HST WFC3/IR FOV. The position of G286.2032+0.1740
is marked with magenta cross.}
%{\bf the color scale should read delta mag}}
\label{fig:var_map}
\end{figure}

HST-WFC3/IR observations of the central cluster region of G286 were
obtained in Cycle 22 and 24 under program IDs 13742 and 14680 (PI:
J. Tan), obtained in  October 2014 and October 2017,
respectively. Observations were carried out in F110W, F160W and F167N
filters and in this study we will focus on the two wide band filters. The
field of view (FOV) for WFC3 is 136\arcsec $\times$123\arcsec, and the pixel
scale is 0.128\arcsec. A 3 $\times$ 3 grid (with 10\arcsec{} overlap
between adjacent pointings) was observed to cover the 6'$\times$ 6'
central region of G286, as shown in \autoref{fig:var_map}.  In both bands three
exposures were obtained for each position in the mosaic with a total
integration time of 897 seconds in F110W and 847 seconds in F160W.
%For
%position 6, marked in Fig. 2, there was one bad frame in both F110W
%and F160W due to enhanced background and these frames were exluded
%from the mosaics. The exposure time is correspondingly shorter.

The data reduction used the STScI processed flt frames and they were
combined using {\tt multidrizzle} with the default parameter
settings. For photometry each tile in the mosaic was handled
individually to avoid potential issues with slight misalignments. The
full width at half-maximum of the point-spread function (PSF) are
0.12\arcsec{} and 0.15\arcsec{} for the F110W and F160W, respectively.

As input for photometry we used the VLT source catalog from
\citet{Andersen17}, which contained 6207 members inside the HST
FOV. The completeness is expected to be
better than 80\% for sources brighter than $K_S$ = 17 and 50\% for
sources down to $K_S$ = 19, as suggested by the artificial star
experiments. 
Aperture photometry was performed with the {\it Daophot} package
in {\it Pyraf}.  For stars located in the overlap regions of different
tiles, we adopted the photometric measurements with smaller errors. An
aperture of 3 pixel radius was used to measure the flux of a source,
and the background was measured in an annulus from 20 to 30 pixels.
%{\bf just mentioning that's a large aperture, but it should be ok given the low surface density of sources, relatively speaking}. 
% yc- the aperture for the flux measurement?
Restricting the list of objects to those with photometric errors
smaller than 0.1 mag in both the F110W and F160W bands results in a list of 5273
sources with photometry in both epochs.

%{Note - In principle there may be some VLT sources further resolved to multiple systems in HST data. So there may not be one-to-one correspondence between HST and VLT detections.}

\subsection{VLT and Gemini observations}
%Our scientific goal in this paper include a brief characterization of the variability of young stars in G286. 

To provide more photometric constraints on an extreme variable star in
this survey (G286.2032+0.1740), we also collected additional
observations including the VLT/HAWK-I JHKs imaging, Gemini/GSAOI Ks
band imaging and Gemini/Flamingos 2 (F2) JHKs imaging.  The VLT
observations were obtained in the programs $087.D-0630(A)$ and
$089.D-0723(A)$ over the period of 2011-2013 (see \citet{Andersen17}
for details). The seeing during the observations was 0.4"-0.6". A
mosaic of 8'$\times$13' was observed. The total exposure times
%across the molecular clump and star clusters 
were 6000s in J, 1500s in H, and 1500s in $\rm K_s$, respectively.  For
this study we used the pipeline reduced and mosaiced images obtained
from each observing block instead of the combined images in
\citet{Andersen17} to be able to follow the time evolution of the
object.
%HAWK-I has a FOV of 7.5' with a pixel scale of 0.106"/pixel. 

%MA, not sure we want this at all but it should not be at this place, even if I placed it here. It should go with the photometry catalogue discussed in 2.1 
%yc- editted

 %$JHK_S$ observations of G286 were obtained with HAWK-I on VLT UT4 under program IDs 087.D-0630 and 089.D-0723 over the period of 2011-2013. HAWK-I has a field of view of 7.5' with a pixel scale of 0.106"/pixel. The seeing during the observations was 0.4"-0.6". A mosaic of 8'$\times$13' was observed. The total exposure times across the molecular clump and star clusters were 6000 s in J, 1500s in H, and 1500s in $K_s$, respectively. More details of this data set, including reduction, calibration and photometry can be found in Andersen et al. (2017). Here we use the catalog in Andersen et al. (2017), which includes 25,875 sources (20,989 with photometric error less than 0.1 mag in each band). Sources down to $K_S$ = 21 mag are detected with a formal error of 0.1 mag. The photometry across the cluster regions is more than 80\% complete for sources brighter than $K_S$ = 17 and 50\% or better complete for sources down to $K_S$ = 19, as suggested by the artificial star experiments.

G286 was observed with Gemini/GSAOI in 2019 March as part of the
proposal GS-2019A-DD-103 (PI: M. Andersen).  GSAOI has a resolution of
0.02\arcsec/pixel and consists of four 2048 $\times$ 2048 pixels
detectors, divided by gaps of $\sim$2\arcsec, providing a total FOV of
almost 85\arcsec$\times$85\arcsec.  Two pointings were obtained, but
here we only discuss the one covering the variable source. A total
exposure time of 45 minutes on-source, was acquired during the run.
The data were reduced in a standard manner using dedicated sky frames
and up to date flat fields, using the {\it gsaoi} package in the
Gemini pyraf package. Before co-addition of the individual frames,
they were corrected for distortion using the program {\it
  discostu}. All the frames were aligned to the first GSAOI frame and
then average combined using bad pixel masks for the individual frames.
Aperture photometry was performed using the {\it Daophot} package in
{\it Pyraf}. An aperture of 3.5 pixel radius was used to measure the
flux, and the background was measured in an annulus from 20 to 35
pixels.
%Point sources in each of the mosaics were identified using the {\it
%  Pyraf} implementation of {\it Daofind} with a $4\sigma$ detection
%threshold. 
%MA we did not use the list from Andersen 2017a?
%yc - I tried both-- but since we only care about one source here 
%I simply deleted the details about source identification.

%For absolute photometry we compare the photometric results for common detections between VLT and Gemini GASOI. -- Not sure how to better describe this procedure.

Gemini/F2 JHK$_s$ imaging was performed in 2019 June and December (proposal
DT-2019A-129 and GS-2019B-FT-109, PI: Y. Cheng). For each observation,
we obtained a total exposure time of 90 seconds in J, 48 seconds in H
and 60 seconds in Ks, respectively. The raw images were reduced using
the {\it Gemini.F2} package provided in the {\it Pyraf}
environment. The aperture photometry was done following similar
procedures as the GSAOI data.

\begin{figure}[ht!]
\epsscale{1.1}\plotone{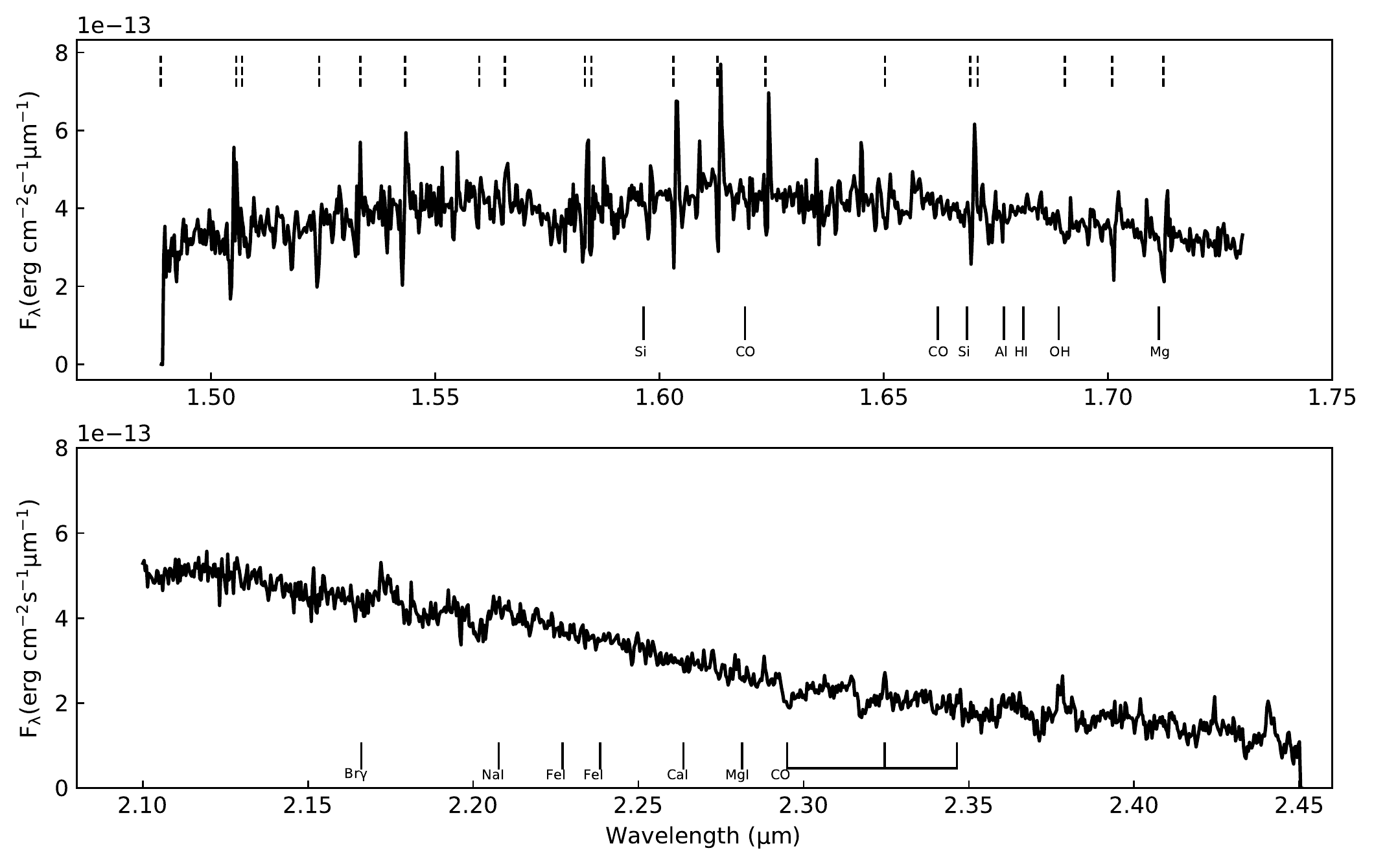}
\caption{
H- and K-band spectra of G286.2032+0.1740. Due to highly variable conditions residuals of OH lines are seen in the H band spectrum, as marked by the dashed black lines. The location of  metallic lines that are seen in absorption for late type stars are marked. The lack of detection of the lines despite the relatively strong continuum suggests veiling.}
% change the Nishiyama+09 reddenning law to Hosek+18, i.e., consistent with the mass estimation.
\label{fig:Var1_spec}
\end{figure}

In 2019 June we also obtained H and K band spectra of G286.2032+0.1740
using F2 under thin cirrus conditions. We used the 2 pixel slit with
the R3K grating resulting in a spectral resolution of 2800 in H and
2900 in K. Ten 120-second exposures were obtained for both the H and
K band spectra in a typical ABBA dither pattern. A telluric star was
observed for both spectral settings. The data were reduced in a
standard manner using flat and Argon lamp observations obtained
after the science exposures. Each science frame was dark subtracted,
flat fielded and sky subtracted using the temporal nearest offset
position before the frames were cross correlated and coadded. The
Argon lamp was used for wavelength calibration. The cirrus did result
in a rather variable sky that has left several OH lines poorly
subtracted in the H band spectrum. These lines are marked in the final
spectrum shown in \autoref{fig:Var1_spec}.
%jct - do a proper reference label for the figure.
% Cheng, did we do any smoothing? }
% No

\section{Results} 

\subsection{Overview of the Region}

\begin{figure}[ht!]
\epsscale{0.8}\plotone{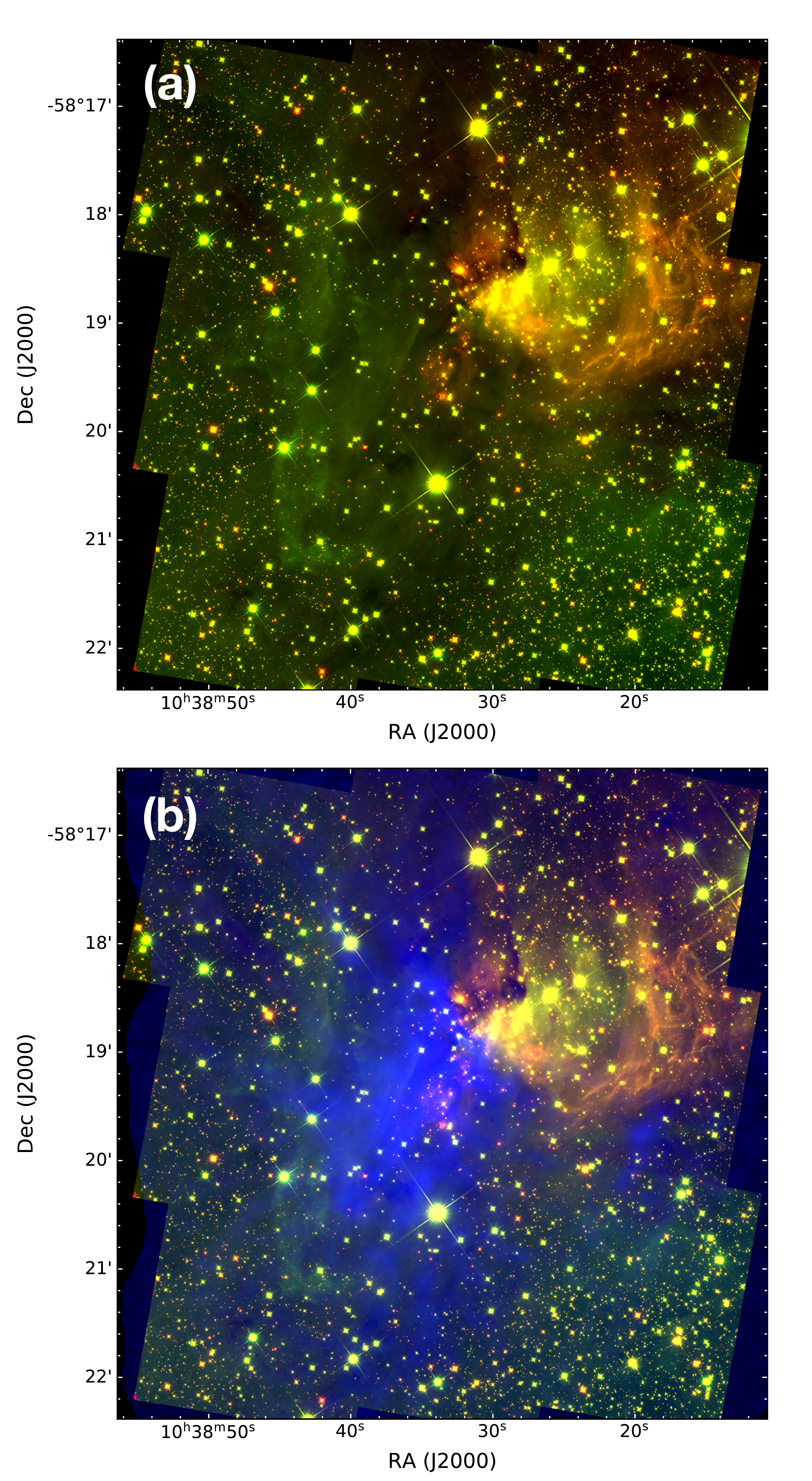}
\caption{
\textit{(a)} HST F110W and F160W (green and red, respectively) color
mosaic of G286. The field of view is 6\arcmin $\times$ 6\arcmin,
corresponding to 4.4~pc $\times$ 4.4~pc for a distance of 2.5
kpc. \textit{(b)} Same as \textit{(a)} but overlaid with the ALMA $\rm
C^{18}O$(2-1) integrated intensity map (from $-$23 to $-$17 km s$^{-1}$) in
blue, which has a spatial resolution of 8.1\arcsec $\times$ 4.8\arcsec
(Cheng Y. et al. 2020, in preparation).  }
%jct - we need to show the high resolution image, so fix this description of resolution
%yc - followoing the AAS reference instructions
\label{fig:overview}
\end{figure}

%jct - do a proper reference label for the figure
\autoref{fig:overview}(a) shows a two-color image of G286 with HST F110W and F160W
data (green and red, respectively).  The stellar component in this
region has been characterized by \citet{Andersen17} with VLT NIR
observations, but the embedded YSO population is better revealed with
our more sensitive and higher resolution HST observations. Some strong
diffuse nebulosity is clearly seen in the northwest, which is
associated with a shell-like HII region, where the stars are less
affected by extinction \citep{Barnes10}. In the central
30\arcsec\ region there is a heavily obscured star cluster (i.e.,
region R1 in \citet{Andersen17}), which appears as redder objects in
this two-color image. Compared with the background/foreground stars
near the edge of the field, there is a relative paucity of stars
extending to the north and south from the center, suggesting existence
of substantial extincting molecular cloud material. This is confirmed
by our ALMA {$\rm C^{18}O$}(2-1) observations (Cheng Y. et al. 2020, in preparation),
%jct - this should be given as a proper reference, i.e., it is submitted.
as shown in \autoref{fig:overview}(b). {$\rm C^{18}O$} is known to be a good tracer of
high column density regions and the integrated emission has a close
correspondence with the dark lanes seen in the HST image.

%jct - make Fig. 1 bigger on the page... 1 vertical column of the 2 figures... fill a whole page as much as possible.

\subsection {Near-IR variability}

\begin{figure}[ht!]
\epsscale{0.8}\plotone{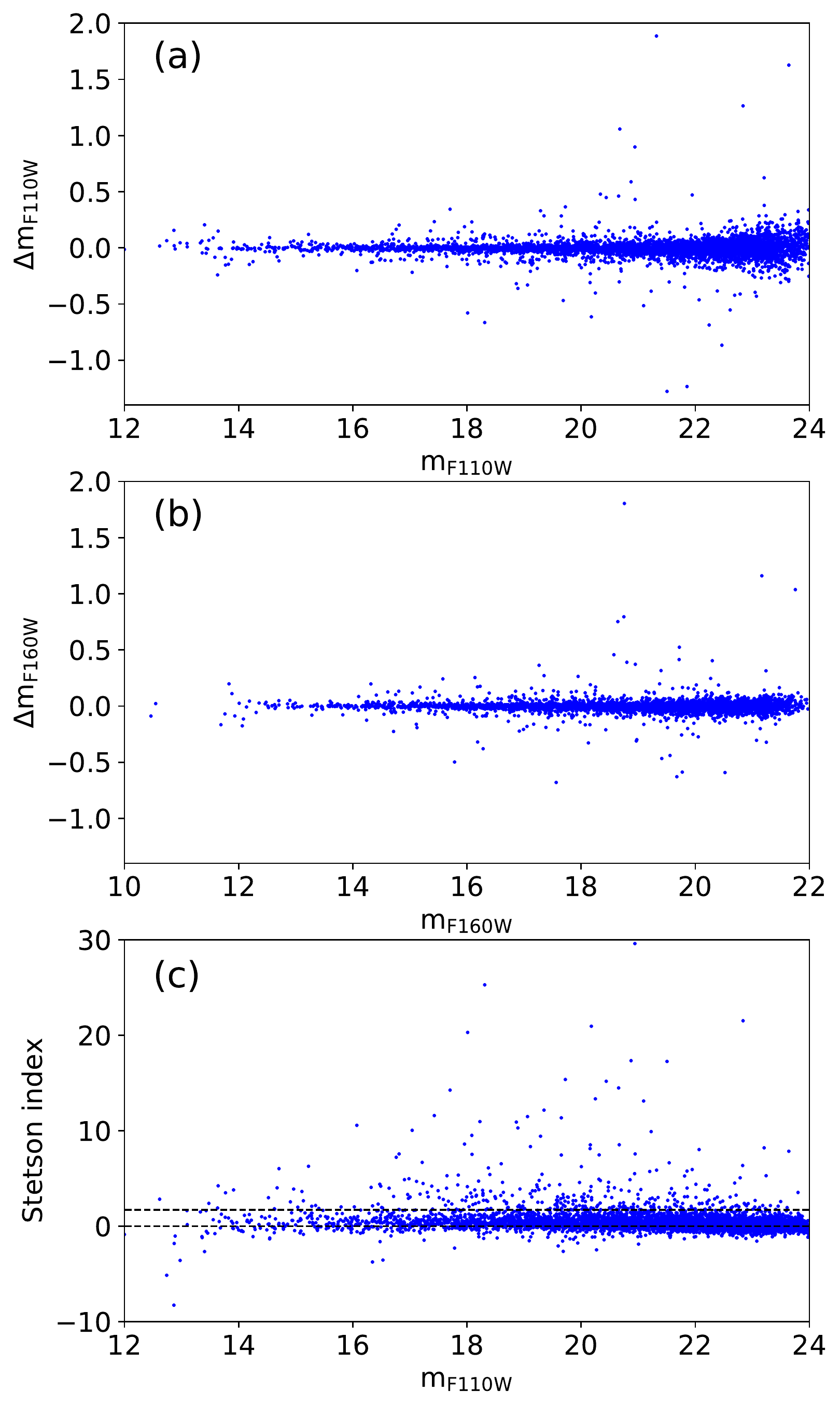}
\caption{
{\it (a)} F110W band variability in two HST epochs against the first
epoch F110W band magnitude. {\it (b)} Same as {\it (a)} but for F160W
band. {\it (c)} Stetson variability index against the first epoch
F110W band magnitude. The dashed lines indicates $S = 0$,
corresponding no variability, and $S = 1.7$, above which we identify as
significant variability. }
\label{fig:var_scatter}
\end{figure}

\autoref{fig:var_scatter}(a) shows the F110W band variation against first epoch F110W
magnitude for the 5273 point sources with photometric errors smaller
than 0.1 mag.  A larger scatter in magnitude variation is seen towards
fainter F110W magnitudes, which is mostly contributed by increasing
photometric uncertainties due to the lower signal for fainter
sources. A Gaussian fit gives a dispersion of $\Delta
m_{\rm F110W} \sim0.03$ for the whole sample. A similar analysis for
F160W band gives a $\Delta m_{\rm F160W} \sim 0.02$ and the
distribution is shown in \autoref{fig:var_scatter}(b).

To quantitatively select stars that are variable, we use the Stetson
variability index \citep[][]{Stetson96}, which is defined as
\begin{equation}
    S = \sum_{i=1}^{p} sgn(P_i)\sqrt{| P_i |},
\end{equation}
where $p$ is the number of pairs of simultaneous observations of an
object. $P_i=\delta_{j(i)}\delta_{k(i)}$ is the product of the
relative error of two observations, which is defined as
\begin{equation}
    \delta_i = \sqrt{\frac{n}{n-1}}\frac{m_i-\overline m}{\sigma_i}
\end{equation}
for a given band. Here $n$ is the number of measurements used to
determine the mean magnitude $\overline{m}$ and $\sigma_i$ is the
photometric uncertainty. The Stetson statistic has been widely used to
characterize variability in multi-wavelength observations
\citep[e.g.,][]{Carpenter01,Rice12}. Since it accounts for the
correlated changes in multiband magnitudes, the Stetson index can be
used to identify variables with relatively low variability compared
with photometric errors.
%On the other hand, the signatures of variability occurring only in a single band will be diluted.  

\autoref{fig:var_scatter}(c) shows the Stetson statistics as a function of F110W
magnitude. For random noise, the Stetson index should be scattered
around zero, and larger positive values indicate correlated
variabilities. An outlier-clipped gaussian fitting of the Stetson
index distribution gives a mean value of $S = 0.2$ and a dispersion of
0.5. Therefore, objects with $S\geq2$ can be considered $3\sigma$
variables and we use $S\geq1.7$ as our criterion for variability
hereafter. Of all the 5273 objects, we have found that 363 (7\%) are
variable. The spatial distribution of these variables is illustrated
in \autoref{fig:var_map}.

The sample consists of heterogeneous populations, including foreground
and background field stars and cluster members. To characterize the
variability for young stars that possess disks, which are mostly  cluster members, we plot $J-H$ versus $H-K$ diagrams in \autoref{fig:color-color}
using the VLT JHKs photometry using the catalog from
\citet{Andersen17}. This color-color diagram is an effective tool to
identify objects with warm circumstellar disks
\citep[e.g.,][]{Meyer97}. Following \citet{Andersen17}, the sample has
two distinct populations: a bluer population ($J-H \approx0.6$), which
is mostly field stars in the foreground of the clump, and a redder
population ($J-H \approx2$) consisting of the cluster content and also
some field star contamination. To detect optically thick disks, we use
the NIR excess criterion devoloped by \citet{Lada92}. We consider
stars to have a NIR excess consistent with an optically thick disk if
they are located to the right of the reddening vector from the M6 main
sequence colors, as shown in \autoref{fig:color-color}. The objects also have to be above
the empirically derived dereddened T-Tauri locus \citep{Meyer97}. In
addition, all objects with a color $J-H<1$ are ignored since they are
expected to be foreground objects. These criteria yield 562 disk
excess candidates, of which 80 are variable at a significant level
($S>1.7$).

The fraction of variables in our identified YSOs that show evidence
for a circumstellar disk is relatively low (14\%) compared with other
NIR surveys \citep[e.g.,][]{Carpenter01,Scholz12,Rice12}, in which
most YSOs have been observed to show a low level of NIR variability,
with typical $K$ band amplitude of $\sim$ 0.15 mag. This is mainly due
to the distance to the cluster. We have increasing photometric errors
for fainter objects (e.g., $\sigma_{\rm F110W} \gtrsim 0.05$ for
$m_{\rm F110W}>22$) and hence it is difficult to detect variability at
a significant level for these faint objects, assuming a typical
variation of 0.15 mag. A higher variable fraction is achieved with a
brightness cut. For example, the variable fraction is 57\% (24/42) for
disk candidates with $m_{\rm F110W}<19$. Furthermore, we only have HST
observations over two epochs separated by 3 years, which may miss some
short-term (weeks to months) variability.

Our observational setup is more suitable to survey long-term, large
amplitude variations. Typical short-term NIR variations, arising from
rotation, hot spots or inner disk inhomogenities, are in the range of
0.1-0.6 mag \citep[][and references therein]{Scholz13}. Larger
amplitude variations in YSOs are usually associated with accretion
outbursts or extinction events. Of all the 5273 objects, 12 have $\Delta m_{\rm F110W}>0.6$
and 7 have $\Delta m_{\rm F160W}>0.6$. The maximum amplitude in F110W
and F160W band are 1.89$\pm$0.03 and 1.80$\pm$0.02, respectively. 
%MA we need error bars for those two numbers; and through this section for the other magnitudes
Of all the 562 YSO
candidates with evidence for a circumstellar disk, 3 (0.5\%) have
$\Delta m_{\rm F110W}>0.6$ and 1 (0.2\%) has $\Delta m_{\rm
  F160W}>0.6$. To search for eruptive events, we further require a
positive change in luminosity and magnitude variations larger than 0.6
in both bands. This gives 5 candidates, with 1 object also
satisfying the disk excess criteria. Detailed inspection suggests that one of them (G286.2182+0.1507) was affected by a bad spot in the detector in the first epoch and hence is excluded in the following analysis. 
%These outburst candidates need confirmation with
%future spectroscopic follow-up.

\begin{figure}[ht!]
\epsscale{1.0}\plotone{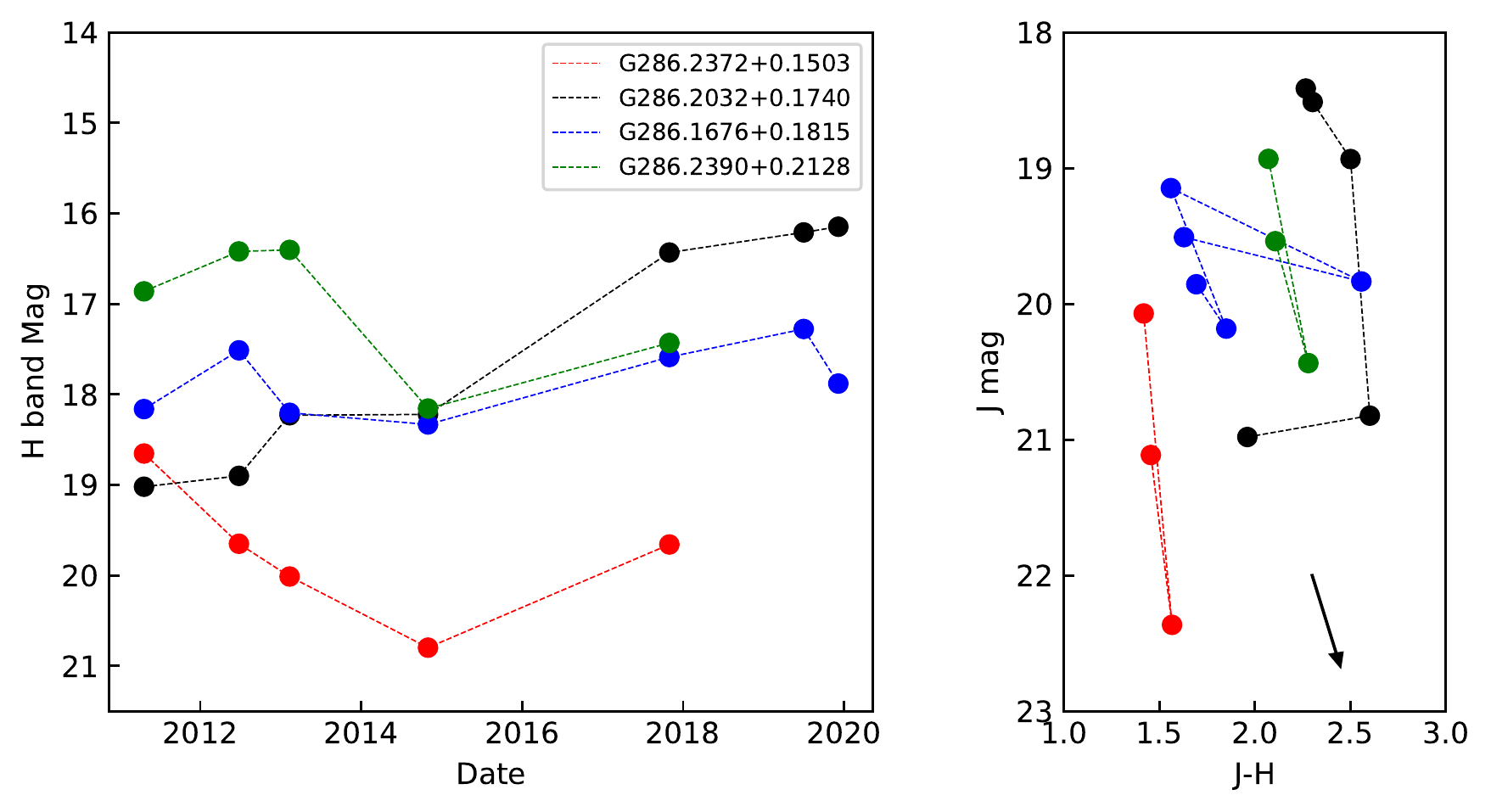}
\caption{
{\it Left:} H band Light curves for the four high amplitude variables. The photometric uncertainties are $<$ 0.1 and not shown here. {\it Right:} Color-magnitude diagram. Note that for epoch 2012 June and 2013 February only H band photometry is available so no data is plotted in this diagram. An extinction vector with $A_K=0.2$ is overplotted using the reddening law of \citet{Nishiyama09}.
}
\label{fig:Vars}
\end{figure}

To investigate the nature of these objects, we have collected more observations, including our early VLT
HAWK-I observations (2011-2013) and recent Gemini GSAOI $Ks$ band
imaging (March 2019) and F2 $JHKs$ band imaging (June 2019, December
2019).  For direct comparison the HST F110W/F160W photometry was
converted into the 2MASS system (i.e., corresponding to $J/H$ bands)
following similar procedures as in \citet{Andersen17b}. In \autoref{fig:Vars} we show light curves and color-magnitude diagrams of the four high amplitude variables in H band, for which we have better sampling. Three of them (G286.2372+0.1503, G286.1676+0.1815, G286.2390+0.2128) show a declining trend from 2012 to 2015, so the brightening between two HST epochs could be understood as returning to their normal luminosity (after a fading event). It is unclear whether we are observing part of a periodic variation or an isolated event. This type of object might be related to either stars going back to quiescent states after an outburst or objects dominated by long-term extinction events similar to the long-lasting fading event in AA Tau \citep{Bouvier13} or some of the faders in \citep{Findeisen13}. The color-magnitude diagram is supportive of the explanation of varying extinction, since most data points of G286.1676+0.1815 and G286.2390+0.2128 seem to follow the direction of the reddening vector. G286.2372+0.1503 has a steeper slope in the color-magnitude diagram, with significant variation in brightness but relative stable color, and hence its variability may also be contributed by other mechanisms besides extinction.     
The other object (G286.2032+0.1740) is the only one of these four objects that exhibits continuous brightening over the observation period of $\sim 8$ years, and thus is more likely to be a long-period accretion outburst event. We discuss its nature further in the following section.

\subsection{An object with extreme variability}

\begin{figure}[ht!]
\epsscale{1.0}\plotone{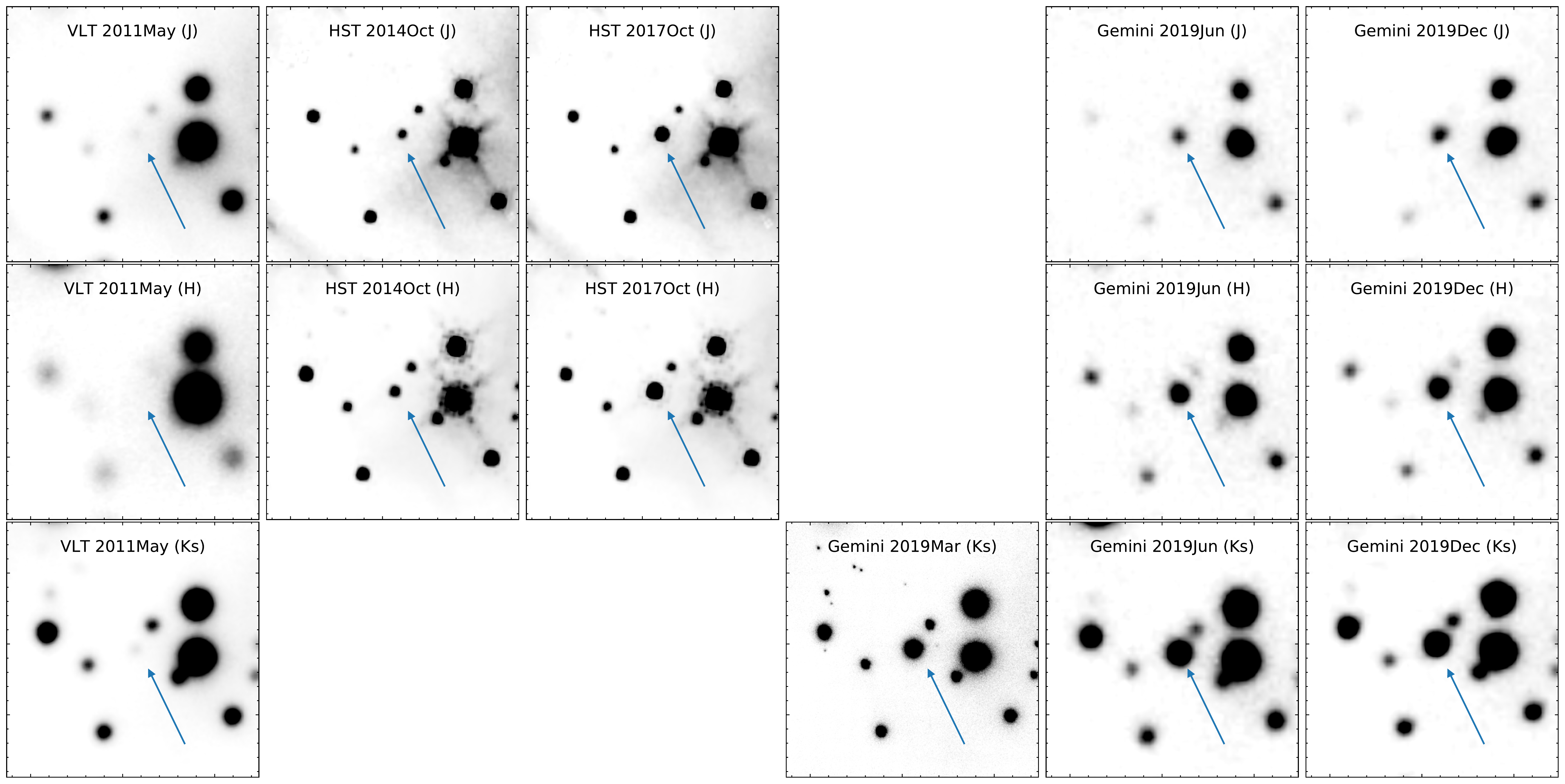}
\caption{
A composite of G286.2032+0.1740 images taken in different bands and at
different epochs. The filters are $J$, $H$ and $Ks$ from top to
bottom, respectively. From left to right are the
$J$(F110W)/$H$(F160W)/$Ks$ images with VLT HAWK-I in 2011 May, F110W/F160W
images with HST in 2014 October, $J$/$H$ images with HST in 2017
October, $Ks$ images with Gemini GSAOI in 2019 March, and $J$/$H$/$Ks$
images with Gemini F2 in 2019 June and December. }
\label{fig:Var1}
\end{figure}

In our variability analysis, we have identified an object with
eruptive variability, i.e., G286.2032+0.1740, located at
($\alpha_{J2000}=10^{h}38^{m}31^{s}.44$, $\delta_{J2000}$ =
$-58^{\circ}$18\arcmin48.2\arcsec).
%MA I think this sentence can be deleted now there is the discussion in section 3.2
G286.2032+0.1740 has the most
extreme variations in both bands, with an brightening of
$\Delta J$ = 1.89 and $\Delta H$=1.79. Further
literature research indicates this object was a faint, virtually
unstudied star prior to the onset of its eruption. 
G286.2032+0.1740 was not previously detected in early NIR surveys such as 2MASS, DENIS and WISE, due to its faintness before eruption.
In \autoref{fig:Var1}, we
show the pre- and post-outburst images of G286.2032+0.1740 in $J$ (top),
$H$ (middle) and $Ks$ band (bottom), taken at different dates, which
clearly reveals a brightness increase in all three bands. The
corresponding lightcurves and photometry are shown in \autoref{fig:Var1_lc} and 
\autoref{table:phot}, respectively. The most striking
contrast is seen in the $Ks$ band: comparing the Gemini GSAOI results
(2019) with the earliest VLT photometry (2011), we measure an
amplitude change of $\Delta Ks$ = 3.5 mag, i.e., a flux increase by a
factor of 25.

Following the light curve morphology categorization in the VVV survey
\citep{Pena17}, G286.2032+0.1740 falls in the ``eruptive variability''
category. In the $H$ band, for which we have better sampling of the
light curve, G286.2032+0.1740 exhibits a monotonic rise over 8 years,
with $H = 19.02$ in May 2011 increasing to $H = 16.11$ in Dec. 2019,
though a lower level scatter is present from 2013 to 2014. On the
other hand, the $J$ band luminosity remains roughly constant until
2015 ($J = 21.14$ in May 2011 and $J = 20.82$ in Oct 2014), and rises
to $J = 18.38$ in Dec 2019. Judging from the $H$ band light curve,
G286.2032+0.1740 went into outburst no later than June 2012, but we
caution that this estimate may be affected by our relatively sparse
sampling of the light curve. G286.2032+0.1740 appears only slightly
brightened from June to December in 2019 and it is not clear if it has
reached the plateau phase.
%{M- \bf last part of sentence would need a little rewrite when the last epoch data is included}.
%A small decrement in Ks band of about 0.1 mag is seen from March to June in 2019.

\autoref{fig:Var1_spec} shows the H and K band spectra taken during June 2019, when
G286.2032+0.1740 was in its bright state. There is a hint of
shallow CO absorption at 2.29 $\mu$m.  The location of the most
prominent lines expected for a late-type star are
%several other common lines are 
marked but there is no clear evidence for emission or absorption
lines,
%suggesting the object is strongly veiled. 
%We have marked the most prominent lines expected for a late-type
%star but
perhaps indicating that they are masked by veiling from the
disk. There is no sign of Br$_\gamma$ emission either, suggesting the
accretion disk may extend all the way to the stellar surface during
this outburst.  We discuss the nature of G286.2032+0.1740 in the next
section.
%This is similar to what was observed for OO serpentis (Kospal et al. 2007). 

\section{Discussion} 

Although the fraction of eruptive variables is very low among YSOs,
they could provide unique insights into specific important processes
occurring in the vicinity of the star, i.e., the star-disk interface,
the inner disk as well as spatial scales beyond 1 AU, depending on
specific mechanisms \citep{Audard14}. A commonly accepted picture is
that these objects are undergoing accretion outbursts, during which
the accretion rate rapidly increases by several orders of magnitude. A
significant fraction of the mass of the star may be accreted in such
bursts. The eruptive YSOs have been traditionally divided into two
classes: FUors, which have large flux increases and long outburst
durations (tens to hundreds of years)\citep[e.g.,][]{Herbig77,Hartmann96}; and EXors, which have recurrent
short-lived outbursts (weeks to months)\citep[e.g.,][]{Herbig89,Herbig08}. Episodic accretion has
several key implications for star formation and evolution, including
solving the ``luminosity problem'' for embedded sources
\citep{Kenyon90,Evans09} and contributing to the luminosity spread of
young star clusters in the Hertzsprung-Russell (HR) diagram
\citep[e.g.,][]{Baraffe09}.
%Additionally, changes in the accretion rate through circumstellar disks could have a profound effect on the merging planetary systems \citep[e.g.,][]{Hubbard17}.  
%There are growing evidence suggesting that these two groups are part of a continuum of outbursts with a range of accretion rates (e.g., Contreras Pena 2017a,b). 

However, many aspects of eruptive YSOs,
including the recurrence time-scale and its relation with
evolutionary stage, are still under active debate\citep[e.g.,][]{Scholz13,Fischer19,Pena19}, partly due to limited numbers of examples.
Both
FUor and EXors categories have fewer than 20 that are known in total
\citep{Reipurth10,Connelley18}.

\begin{figure}[ht!]
\epsscale{0.6}\plotone{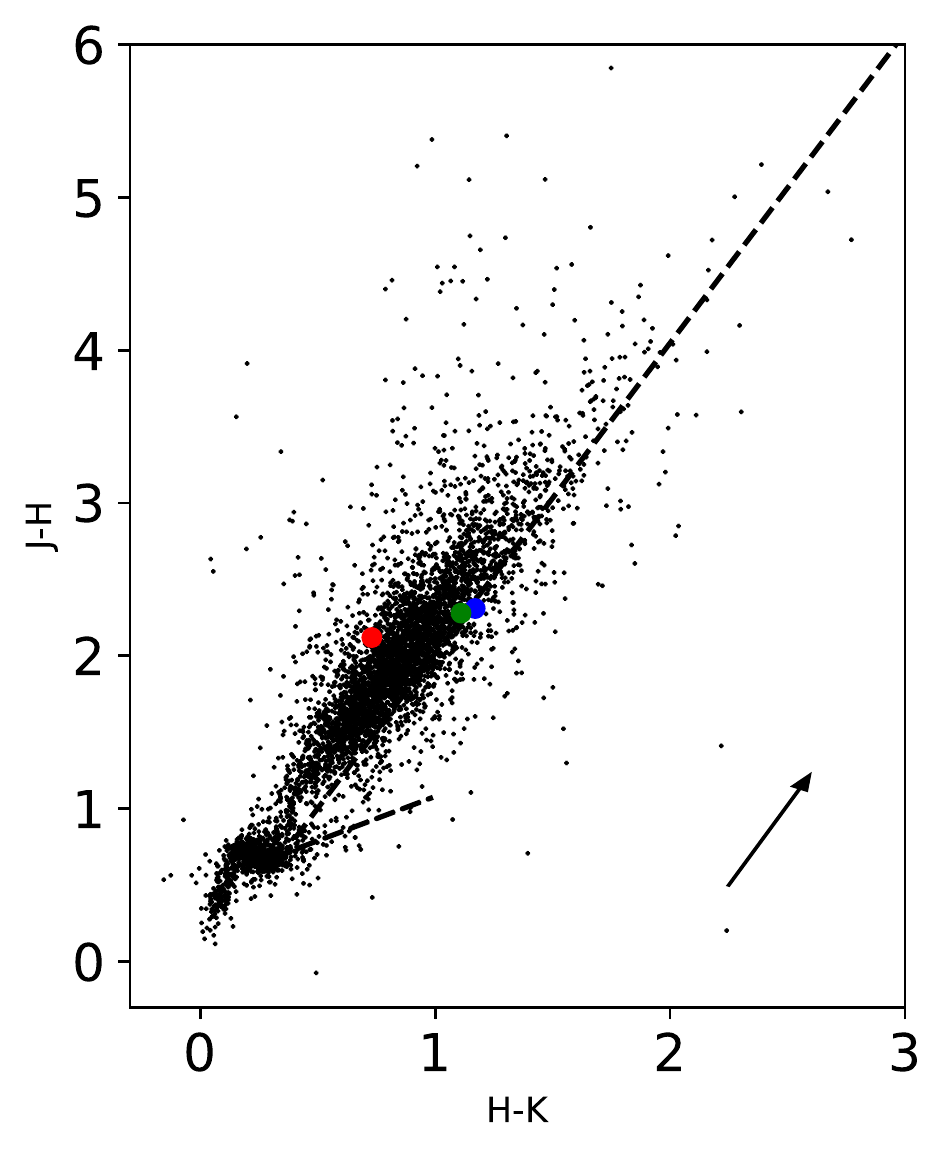}
\caption{
$J-H$ vs. $H-K$ color-color diagram. Overplotted are
%the main-sequence dwarf sequence (Bessell \& Brett 1988), 
the reddening vector extending from an M6 spectral type and the
T-tauri locus from \citet{Meyer97}. The red, blue and green dots
denote G286.2032+0.1740 at three epochs, i.e., 2011 May, 2019 June
and 2019 December, respectively. An extinction vector with $A_K=0.5$
is overplotted, using the reddening law of \citet{Nishiyama09}. }
%{\bf if you could include a reference to the reddening vector as well as indicating with an arrow Ak=1 or similar}}
\label{fig:color-color}
\end{figure}

%The occurrence rate and duty cycle of FUor events is crucial to establish, given our poor understanding of accretion histories during star formation and pre-main-sequence evolution. However, FUor outbursts are rare. Over the past seven decades, fewer than 13 actual FUor outbursts have been recorded, with another ∼13 sources identified as FU Ori-like based on their present spectra and spectral energy distributions in a hypothetical post-outburst state (Reipurth \& Aspin 2010; Connelley \& Reipurth 2018). Therefore, very few FUor objects are well observed and studied during their outbursts and hence our knowledge of the progenitors, the range in light-curve rise shapes, and the early outburst spectroscopic characteristics is severely limited.
%Since G286.2032+0.1740 has not been well studied and categorized in previous work, we need more careful treatment to establish its nature as an YSO. 
Comparing with known eruptive variables classes, G286.2032+0.1740,
characterized by a long-term large-amplitude rising light curve,
resembles an FUor object in its temporal behavior. In principle,
high-amplitude variability in the NIR can be produced by various
physical phenomena, including evolved giant and supergiant stars like
Mira variables, cataclysmic variables and active galactic nuclei (AGN), etc
\citep[see][for a discussion]{Catelan13}. However, none of these
possibilities is consistent with the characteristics of
G286.2032+0.1740, including its faintness, NIR color and shape of the light curve. 
For example, there is no indication of periodicity from the light curve
of G286.2032+0.1740, in contrast with what is expected for evolved stars
like asymptotic giant branch (AGB) stars. The slowly rising light curve over years
is also inconsistent with a nova outburst event\citep{Warner03}. The NIR
variability of AGN, on the other hand, is relatively smooth, but with smaller 
amplitude \citep{Enya02,Cioni13}. Furthermore, the fact that G286.2032+0.1704 is 
located in the Galactic plane with moderate extinction also makes it highly unlikely
to be a background object like AGB star or AGN.
%MA I would suggest to include the fact that it's in the plane and the moderate extinction and close to the core makes it unlikely it is a background object. Perhaps also that it is a point source in the HST images? 
In \autoref{fig:color-color} we overplot the $J-H$ vs. $H-K$ colors of
G286.2032+0.1740 for three epochs with sufficient data (i.e., where
the different bands were obtained close to each other within 5 days)
(2011 May, 2019 June, 2019 December).
% now we should have three data points (i.e., plus the 2019 Dec one), will fix this in the next hour
At the more recent two epochs G286.2032+0.1740 appeared close to the
boundary of the disk excess criterion, while in the early epoch
G286.2032+0.1740 was to the left of that boundary, indicating a later
evolutionary stage without much disk/envelope material. However, $JHK$
observations are known to be less sensitive to disks around young
stars, compared with $L$-band observations or mid-infrared diagnostics
\citep[e.g.,][]{Haisch00} and stars with disks may drift in $JHK$
color space, rendering a smaller detection rate with only single epoch
observations \citep{Rice12}. Overall, given its location in a known
active star-forming region and its photometric behavior, we consider
that G286.2032+0.1740 is more likely to be an outbursting YSO.
Similar to other FUor/FUor candidates, G286.2032+0.1740 has a spectrum
lacking emission lines.
%, although the relatively low signal to
%noise given the faint nature of the source may hamper such detections.
%It's not really low SNR, it's just without features. 
This relatively featureless spectrum, as well as possible CO absorption, is broadly consistent with the FUor category. A similar example is VVVv721 \citep{Pena17b,Guo20}, which is classified as a FUor and characterized by having CO absorption and broad $\rm H_2O$ absorption bands, with a lack of other photospheric features. 
In the case of G286.2032+0.1740 some doubts will remain since the CO absorption features are very weak compared to typical FUors \citep{Reipurth10} and some other common characteristics of FUors, like broad band water vapor absorption, are also not clearly seen.
G286.2032+0.1740 has a relatively slow rise in its light curve (rise time $>$ 8 years), which is similar to the classical FUor V1515 Cyg \citep{Kenyon91}and VVVv721\citep{Pena17b}. This slow rise may be explained as resulting from thermal instabilities that spread from the inner regions towards the outer parts of the accretion disc \citep[see e.g.,][]{Audard14}.

%The relatively featureless spectrum makes it difficult to
%dentify the nature of the source, 
%e.g., we are unable to find signatures
%expected for typical FUor or EXors \citep[e.g.,][]{Connelley18}.

\begin{figure}[ht!]
\epsscale{1.}\plotone{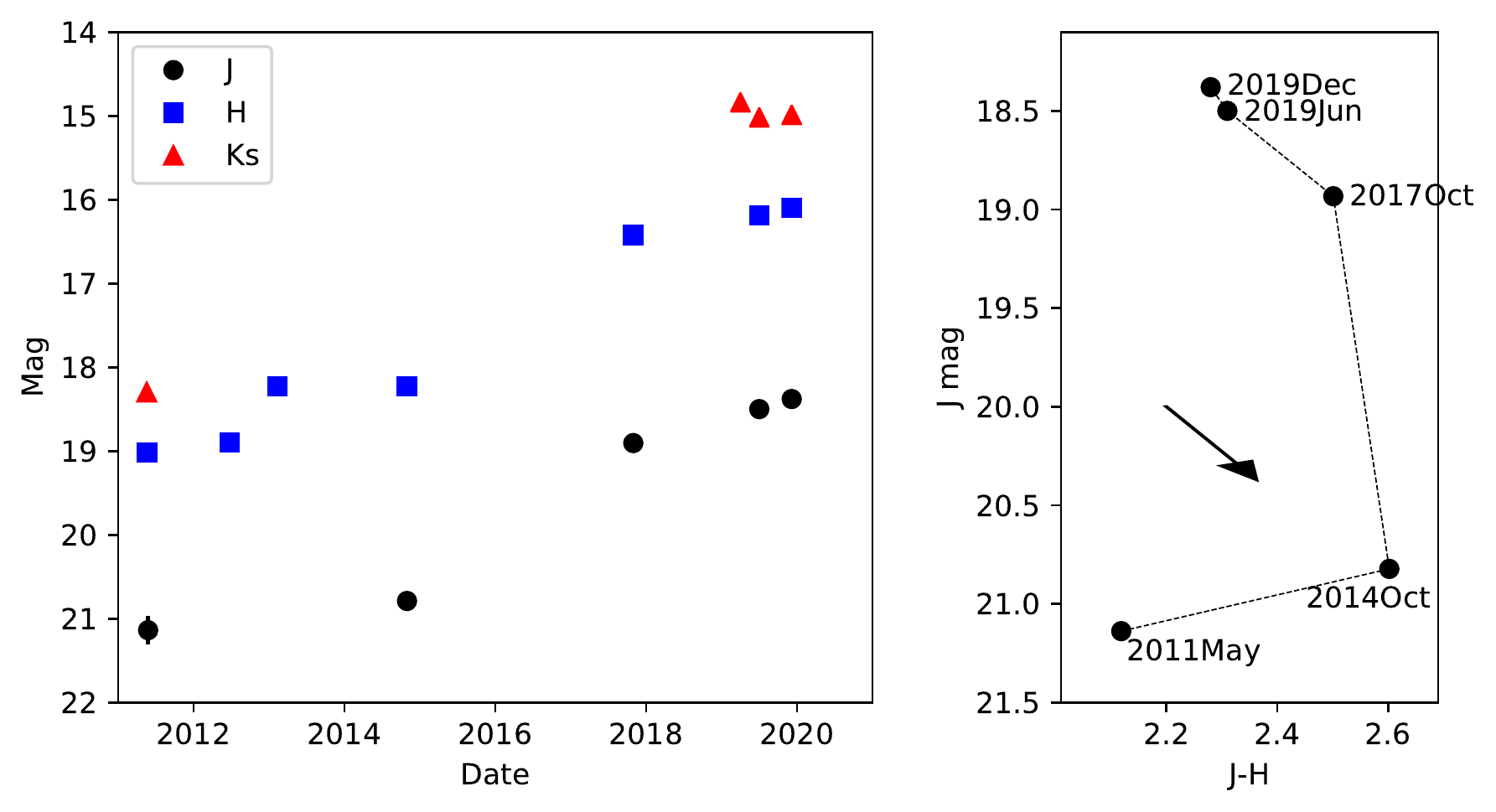}
\caption{
{\it Left:} Light curve for G286.2032+0.1740 over 8 years. $J$, $H$,
$Ks$ data are represented as circles, squares and triangles,
respectively. {\it Right:} Color-magnitude diagram for
G286.2032+0.1740. An extinction vector with $A_K=0.1$ is overplotted,
using the reddening law of \citet{Nishiyama09}.}
\label{fig:Var1_lc}
% change the Nishiyama+09 reddenning law to Hosek+18, i.e., consistent with the mass estimation.
\end{figure}

Apart from accretion bursts, variable extinction may also be the
reason for some extreme variation cases. For example, the variability
of ESO-Oph-50 is explained by a low mass star seen through
circumstellar matter, with changing inhomogeneities in the inner parts
of the disk \citep{Scholz15}. In the bright state the emission is
consistent with a photosphere reddened by circumstellar dust, while in
the faint state we are observing bluer scattered light since the
direct stellar emission is blocked. However, the color behavior of
G286.2032+0.1740 is inconsistent with this scenario. In \autoref{fig:Var1_lc} we plot
the color-magnitude diagram of G286.2032+0.1740. The trajectory can be
divided into two stages: from May 2011 to Oct 2014 G286.2032+0.1740
became redder with slight brigtenning in the $H$ band. In the second
stage, G286.2032+0.1740 turned bluer and brighter, which is in
contrast to ESO-Oph-50 (bluer when fainter), but consistent with some
outburst cases \citep[e.g.,][]{Aspin09}. The magnitude changes in this
stage are also
%jct - check this word:
steeper
%faster
than expected from the reddening vector and thus cannot be attributed
to variable extinction. The color variation in the first stage (from
2011 to 2014) is more erratic, in which G286.2032+0.1740 was reddened
by 0.5 mag but kept similar $J$ band brightness. Unfortunately we only
have a handful of data points in this pre/early outburst stage and
thus cannot give more constraint on the nature of its color variation.

In the earliest epoch (2011), G286.2032+0.1740 was a faint object with
a $K$ magnitude of 18.3, suggesting its nature as a very low-mass YSO
and/or it was observed through substantial extinction. To
quantitatively estimate its mass, we compare its $JHK$ photometry with
the predictions from the \citet{Baraffe15} isochrone, assuming a
typical age of 1~Myr for G286 \citep{Andersen17}. Depending on which
two colors are used for de-reddening, we obtain a range of masses from
0.06 to 0.10~$M_\odot$. The mass estimation falls in a similar range
(0.05 to 0.12~$M_\odot$) with varying assumed ages from 0.5 Myr to 2
Myr. Note that even the earliest epoch data here may not represent the
pre-outburst quiescent state, so this estimation should be considered
as an upper limit. If this is confirmed by further spectroscopic
observations, G286.2032+0.1740 provides a unique case to study the
extreme variability for YSOs in the very low-mass regime, for which
our knowledge is still sparse. Very low-mass stars and brown dwarfs
have been observed to have both low-level periodic variability, and
more irregular high amplitude variability, but typically only with
I-band amplitude changes up to 1 mag \citep{Scholz05,Bozhinova16}. In
terms of high amplitude variables ($>$ 3 mag) that are likely
associated with strong accretion outbursts, there are very few cases
reported in the very low-mass range ($<0.5\:M_\odot$) \citep[e.g.,
  ASASSN-13db, (CTF93)216-2,][]{Holoien14,
  Caratti11}. G286.2032+0.1740, with mass $<0.12\:M_\odot$, is the
lowest mass YSO with a strong outburst found so far. Combining its
very low mass and strong outburst, G286.2032+0.1740 is apparently an
extreme case of YSO varability. Since the object is currently near its
brightest state, it gives a unique chance to characterize a very low
mass YSO in its eruption stage.

%{\bf MA: we might want to include a few sentences on speculations that
%  disks survive shorter in massive star clusters and thus objects like
%  this one would be less common in such an environment.  }

%Considering a typical field inital mass function most stars have masses $<$ 0.5 M$_\odot$, however, our knowledge of outbursts in more typical low-mass YSOs is severly limited since there are very few outburst cases in that stellar mass range reported \citep[e.g., ASASSN-13db, (CTF93)216-2,][]{Holoien14, Caratti11}. 

\begin{deluxetable}{lcccc}
\tabletypesize{\scriptsize}
\caption{Photometry of G286.2032+0.1740}
\label{table:phot}
\tablehead{
\colhead{Date(y.m)} & \colhead{Instrument} & \colhead{J} & \colhead{H} & \colhead{Ks}
}
\startdata
2011.5   & VLT/HAWK-I & 21.14$\pm$0.17 & 19.02$\pm$0.11 & 18.30$\pm$0.08 \\
2012.6   & VLT/HAWK-I & ... & 18.90$\pm$0.06 & ... \\
2013.2   & VLT/HAWK-I & ... & 18.23$\pm$0.09 & ... \\
2014.10   & HST/WFC3 & 20.82$\pm$0.03 & 18.22$\pm$0.02 & ... \\
2017.10   & HST/WFC3 & 18.93$\pm$0.01 & 16.43$\pm$0.01 & ... \\
2019.3 & Gemini/GASOI & ... & ... & 14.84$\pm$0.01\\
2019.6 & Gemini/F2 & 18.50$\pm$0.03 & 16.19$\pm$0.02 & 15.02$\pm$0.01\\
2019.12 & Gemini/F2 & 18.38$\pm$0.03 & 16.11$\pm$0.02 & 14.99$\pm$0.01
\enddata
%\tablenotetext{a}{For velocity dispersion we take the linear average of 4 strips.}
\end{deluxetable}

% will add another column on the plot to include Dec 2019 data.

\vspace{5mm}
\facilities{HST(WFC3), Gemini GS-2019A-DD-103, ESO}


\software{
          }



%\bibliographystyle{aasjournal}
\bibliography{refer}



\end{document}